\documentclass[twocolumn, showpacs, prl]{revtex4}
\usepackage{amsmath,amssymb}
\usepackage{graphicx}
\usepackage{bm}
\usepackage[usenames]{color}

\begin{document}


\title{ Charge Ordering and Ferroelectricity in Half-doped Manganites} 

\author{Kunihiko Yamauchi$^{1}$}
\author{Silvia Picozzi$^{2}$} 
\affiliation{
1. ISIR-SANKEN, Osaka University, 8-1 Mihogaoka, Ibaraki, Osaka, 567-0047, Japan\\
2. Consiglio Nazionale delle Ricerche (CNR-SPIN), 67100 L'Aquila, Italy 
}

\date{\today}

\newcommand{\pscmo}{Pr(Sr$_{0.1}$Ca$_{0.9}$)$_{2}$Mn$_{2}$O$_{7}$}
\newcommand{\pcmo}{PrCa$_{2}$Mn$_{2}$O$_{7}$}
\newcommand{\pcm}{Pr$_{0.5}$Ca$_{0.5}$Mn$_{}$O$_{3}$}

\begin{abstract}
By means of density-functional simulations  for half-doped manganites, such as pseudocubic Pr$_{0.5}$Ca$_{0.5}$Mn$_{}$O$_{3}$ $\:$ and bilayer PrCa$_{2}$Mn$_{2}$O$_{7}$, we discuss the occurrence of ferroelectricity and we explore its crucial relation to the crystal structure and to peculiar charge/spin/orbital ordering effects. In pseudocubic Pr$_{0.5}$Ca$_{0.5}$Mn$_{}$O$_{3}$, ferroelectricity is induced in the Zener polaron type  structure, where  Mn ions are dimerized. In marked contrast, in bilayer PrCa$_{2}$Mn$_{2}$O$_{7}$, it is the displacements of apical oxygens bonded to either Mn$^{3+}$ or Mn$^{4+}$ ions that play a key role in the rising of ferroelectricity. Importantly, local dipoles due to apical oxygens are also intimately linked to charge and orbital ordering patterns in MnO$_2$ planes, which  in turn contribute to polarization.  Finally, an important outcome of our work consists in proposing Born effective charges as a valid mean to quantify charge disproportionation effects, in terms of anisotropy and size of electronic clouds around Mn ions.
\end{abstract}

                              

\maketitle
{\em Introduction.} ``Improper multiferroics''\cite{maxim}, materials where  ferroelectricity is driven by either spin ordering (SO), charge ordering (CO) or orbital ordering (OO), constitute a playground for
the physics of cross-correlation: 
the coupling between the different orderings and structural distortions  is indeed much stronger and richer than  in conventional covalency-driven ferroelectrics.\cite{silvia.ederer} 
Recently, a complex mechanism of ferroelectricity driven by SO and CO has been proposed both in {\it half-doped} manganites and in nickelates, supported by theoretical studies.\cite{efremov, gianluca, ggnickel} 
In such systems, it is considered that slightly charge-diportionated two magnetic ions form dimers via double exchange interaction, resulting in electric dipoles.\cite{khomskii.natmat}
In  the particular case of bilayer-manganite \pscmo, 
it has been experimentally suggested that the transition between two different CO phases is 
accompanied by the rotation of orbital stripes, in turn related to ferroelectricity.\cite{tokunaga}  

This letter is meant to provide insights into cross-correlation phenomena and the ferroelectric instability in half-doped manganites via first-principles approaches. 
Along these lines, there are two delicate problems casted by previous DFT studies: 
{\em i}) 
it is difficult to unambiguously identify the ground state\cite{patterson, gianluca, vincenzo} 
between two different types of atomic (and related electronic) arrangements: {\em a}) centrosymmetric checkerboard (CB) CO pattern of Mn$^{3+}$ ($t^{3}_{2g}$$e^{1}_{g}$) and  Mn$^{4+}$ ($t^{3}_{2g}$$e^{0}_{g}$) ions 
and {\em b}) ferroelectric-active  ``Zener polaron'' (ZP) structure, where the two equivalent Mn$^{3.5+}$ ions are dimerized via a ``bond-centered" charge; {\em ii}) in the CB structure, the ``local charge'' of Mn$^{3+}$ and Mn$^{4+}$ ions is in general ill defined \cite{franceschetti} and a negligible charge separation (in contrast with the nominal valence difference of 1$e$) is often obtained. 
Here, we address these two problems by reporting our DFT results on pseudocubic \pcm\ (PCMO) and  bilayer \pcmo. 

{\em Half-doped manganites: phenomenology.}
In PCMO, as in many half-doped manganites, the MnO$_2$ layer shows the CE-type AFM configuration,  
consisting of  double zigzag FM chains antiferromagnetically coupled both in-plane and out-of-plane, 
which also invokes CO in the form of Mn$^{3+}$ and Mn$^{4+}$ ions, arranged in a checkerboard pattern, 
below $T_{\rm CO}$=245K (much higher than the magnetic transition temperature $T_{\rm N}$=175K).   
Between $T_{\rm N}$ and  $T_{\rm CO}$, there is no long-range magnetic ordering, but likely a persisting spin-fluctuating behaviour; we remark, however, that the system remains insulating in the presence of CO. 
Non-bonding $t_{2g}$ electrons give rise to a localized spin, $S=\frac{3}{2}$, responsible for the AFM super-exchange coupling, whereas $e_{g}$ orbitals, strongly hybridizing with O $2p$ orbitals and consequently  producing broad bands,  are responsible for the double exchange mechanism.
The half-filled $e_{g}^{\uparrow}$ subband is a typical example of the cooperative Jahn-Teller (JT) effect, where the JT distortion of MnO$_{6}$ octahedrons results in  opening of the gap in the $e_{g\uparrow}$ band and in an insulating ground state. 
The crystal structure changes symmetry from 
$Pbnm$ (at room temperature) to lower symmetry in the CO phase, with two possibilities: CB-like centrosymmetric $Pbnm$ (alternatively, $P2_{1}/m$) and ZP-like polar $P2_{1}nm$ at low temperature. 
The stability of these phases and  the ferroelectric/dielectric properties are discussed later. 
We took CD (charge-disordered) $Pbnm$ structure from Ref.\cite{jirak}, 
and 
ZP-CO $P2_{1}nm$  structure from Ref.\cite{daoud-aladine}, consistent with previous DFT studies. \cite{vincenzo}

\begin{figure}[th]
\centerline{\includegraphics[angle=0, width=0.99\columnwidth]{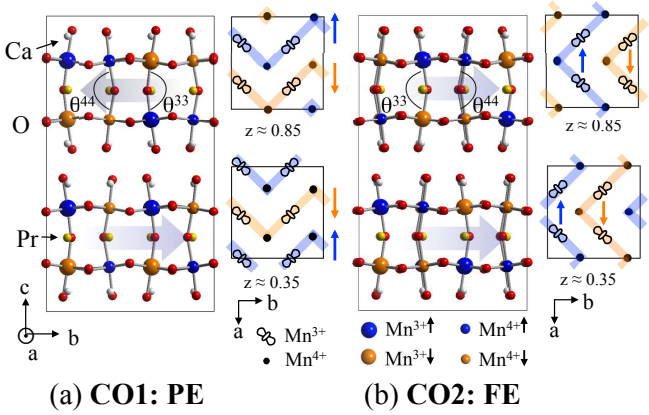}}
\caption{\label{fig:COpattern} 
CO/OO patterns in (a) CO1 phase,  
with CO stripes running along $a$, OO zigzag lines along $b$ 
 and in (b) CO2 phase, with CO stripes running along $b$, OO zigzag lines along $a$, in bilayer \pcmo.
 Inside each bilayer, the inter-layer coupling is AFM. 
The Mn-(apical O)-Mn angle is denoted by $\theta^{33}$ ($\theta^{44}$) between Mn$^{3+}$ (Mn$^{4+}$) ions. 
The blue arrows indicate the direction of electric dipole moment caused by the difference between $\theta^{33}$ and $\theta^{44}$.
}
\vspace{-0.5cm}
\end{figure}

In bilayer-manganite \pscmo $\:$ as well,  the MnO bilayer shows the CE-type AFM configuration,  
with peculiar different CO transitions. 
Upon cooling, the system was shown to undergo two transitions at $T_{\rm CO1}\sim$370 K and at $T_{\rm CO2}\sim$310 K (while $T_{\rm N}\sim$153K), 
where the system shows a transition from CD phase to the ``CO1'' phase, forming Mn-$e_{g}$ orbital chains along the $b$-axis, 
and then to the ``CO2'' phase, forming orbital chains along the $a$-axis (which requires a  90$^{\circ}$ rotation of the OO pattern, cfr Fig.\ref{fig:COpattern}). 
Interestingly, the orbital rotation occurred concomitantly with a rearrangement of Mn$^{3+}$/Mn$^{4+}$ cations, leading to a different stacking of the CO-pattern between the bilayers. 
Although second-harmonic-generation measurements revealed a non-centrosymmetric state\cite{itoh}, 
ferroelectric/pyroelectric current measurements could not be carried out due to leakage problems. 
Though truly remarkable, the study of Tokunaga {\em et al.} leaves many open questions (which will be addressed in this paper), such as how  large the polarization is,  if and how the polarization is coupled to the
SO, CO, and OO, the role of correlation effects vs structural effects, etc. 

\begin{figure}[!ht]
\vspace{0cm}
\centerline{\includegraphics[bb=12 0 360 660, clip, angle=0, width=0.99\columnwidth]{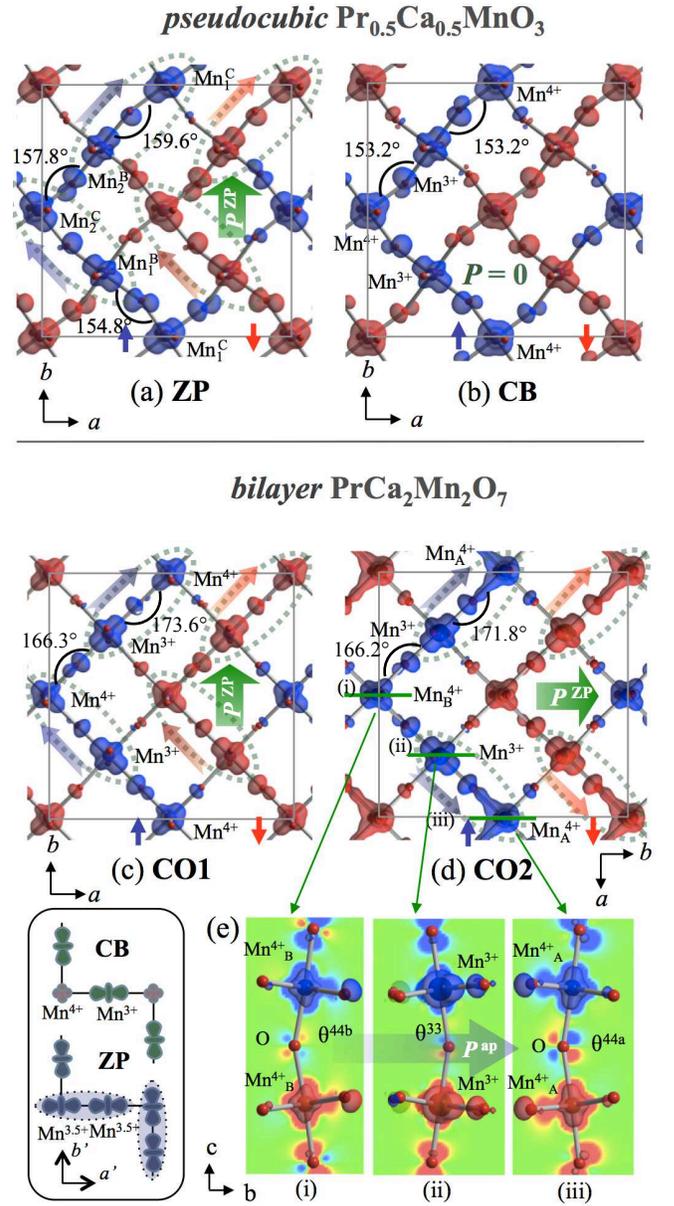}}
\caption{\label{fig:chgsec} 
An isosurface and sections in $ab$ plane 
 of spin density (blue:up, red:down) of Mn-$e_{g}^{1}$ occupied state
 at (a) ZP and (b) CB in PCMO ($z$$\simeq$0) and  
at (c) at CO1 and (d) at CO2 phases in \pcmo\ ($z$$\simeq$0.85). 
Note the $ab$ frame is shifted and rotated for a better comparison.  
The direction of local polarization $P$ and of asymmetric $e_{g}$ electron hopping from Mn$^{3+}$ to Mn$^{4+}$ sites are shown by translucent arrows.
(e) The same isosurface and the section of CO2, cut in the $bc$ plane. 
Inset; Schematic picture of CB and ZP orbital patterns. Dimerized Mn ions are surrounded by dotted ellypses. 
}
\vspace{-0.5cm}
\end{figure}

\bigskip 

{\em Computational details.} DFT simulations were performed using the VASP 5.2 code\cite{vasp} and the PAW pseudopotentials \cite{paw} within the GGA+$U$ formalism\cite{ldau} ($U$=3 eV and $J$=0 eV for Mn $d$-states\cite{Luo}; other values of $U$ are also tested).   
${\bm K}$-point meshes of (3, 3, 2) and (2, 2, 1) were used for the Brillouin zone integration for PCMO and \pcmo. 
%
%
For the A-site 
ordering, 
we consider 
interlayer Ca-Pr-Ca sandwich stacking along the $a$-direction for PCMO (to do this, the $Pbnm$ symmetry is kept) 
and 
Ca-Pr-Ca sandwich along the $c$-direction for \pcmo. The Sr doping was neglected, since it is expected not to be relevant for our discussion.
 {\em i.e.} the one that simultaneously optimizes the energy gain deriving from structural/orbital/spin degrees of freedom.  We succeeded in stabilizing all the CO patterns in the insulating state, even with $U$=0.


{\em Born effective charges and charge disproportionation}. In PCMO,  when considering the CE-AFM configuration and the 
reduction of symmetry from the CD phase into the CB phase, 
 Mn sites split  into Mn$^{3+}$ and Mn$^{4+}$ sites; on the other hand, 
 in the ZP phase,  the two
original Mn sites (Mn$_{1}$ and  Mn$_{2}$), each split into corner and bridge sites along the spin chain (e.g. Mn$_{1}^{\rm C}$ and Mn$_{1}^{\rm B}$), where Mn$_{1}$ and Mn$_{2}$ ions are dimerized. 
When comparing the local charge of two different Mn sites, obtained by integrating the charge density within atomic spheres, the charge separation is smaller than 0.01$e$, whereas we quantified  difference of the local spin moment between Mn$^{3+}$ and Mn$^{4+}$ ions as 
0.04$\mu_{\rm B}$ and 0.33$\mu_{\rm B}$ in the ZP and CB phases, respectively. 
In both cases, such tiny differences (in contrast to the nominal values of 1$e$) are affected by large uncertainties, since they depend on the choice of the input radii within which the charge/spin density is integrated. 
Due to the strong double-exchange nature of Mn-$e_{g}$ electrons \cite{PCMO_Anisimov},  the
Mn-$d$ charge is more like  ``bond-centered" rather than ``site-centered", so that the on-site charge based on an ionic picture is ill defined. 
Luo {\it et al.} proposed the Mn-$d$ {\it orbital occupancy} as an appropriate indicator of charge separation, giving $\Delta n$=0.17$e$ in electron-doped CaMnO$_{3}$.\cite{Luo}  
Actually, the difference between Mn$^{3+}$ and Mn$^{4+}$ sites in the CB phase is clearly seen in the occupied orbital shape, shown in Fig.\ref{fig:chgsec}. 
We go one step further by observing that the topology (size and anisotropy) of the orbitals should be described by  ``tensorial "instead of ``scalar"  quantities. 
Therefore, here we introduce the Born effective charge (BEC), $Z^{*}_{\alpha \beta}=\delta P_{\alpha}/ \delta u_{\beta}$ ($\alpha, \beta = x,y,z$) for each ionic displacement $u$. 
The diagonalization of $Z^{*}$ leads to eigenvalues $Z^{\rm d}_{x', y', z'}$ and eigenvectors, 
which correspond to the direction of the response of charge density around an ionic site upon displacement. 
This represents a practical way to characterize  the anisotropic nature of $pd$ hybridization along Mn-O bonds. 
For example,  the Mn$^{3+}$ ion in CB phase shows \\
$Z^{*}_{\rm Mn^{3+}}=
{
\footnotesize
\begin{pmatrix}
   {3.67} &  {2.94} &  0.61 \\ 
  {2.38} &   {4.89} &  1.48 \\ 
  -0.51 &  -0.27 &   4.53 \\ 
\end{pmatrix}
}
$ 
with  
$Z^{\rm d}_{\rm Mn^{3+}}=
{\footnotesize
\begin{pmatrix}
 7.91\\ 
 1.20\\
 3.91 \\ 
\end{pmatrix},   } $
where the first (second) component of $Z^{\rm d}$ is parallel (perpendicular) to the $3z^{2}$-$r^{2}$ orbital lobe in MnO$_2$ planes and the third component parallel to the $c$ axis (Fig.\ref{fig:chgsec} inset). 
This clearly reflects the bonding nature of Mn $e_{g}$ electrons along the spin chain, consistent with the plot of the spin-density (up- minus down-spin components of charge density) reported in Fig.\ref{fig:chgsec}; the BEC is enhanced along the direction where Mn states shows strong $pd$ hybridization with neighbouring O $p$-states. 
On the other hand, the Mn$^{4+}$ ion in the CB phase shows   more
isotropic $Z_{d}$ components almost parallel to the ($a$, $b$, $c$) axes  (cfr Tab.\ref{tbl:bec.pcmo}).
In the ZP phase, we got  anisotropic (i.e. Mn$^{3+}$-type)  $Z^{\rm d}$ both at Mn$_{1}$ and Mn$_{2}$ sites. 

\begin{table}[ht] 
\vspace{-0.5cm}
\caption{
Diagonalized Born effective charge $Z^{\rm d}$ ($e$)  at Mn sites. 
Subscripts denote the direction; $a'$ and $b'$ are parallel to diagonal axis, $(a\pm b)/2$, as bonding and non-bonding direction, respectively. } 
\label{tbl:bec.pcmo}
%
\begin{tabular}{cc}
%
\begin{tabular}[t]{cccc}
\multicolumn{4}{l}{PCMO}\\

\toprule
\multicolumn{4}{c}{CB}\\
  \colrule
Mn$^{3+}$ & 
(       7.9$_{a'}$ &
        1.2$_{b'}$ &
        3.9$_c$ ) \\
Mn$^{4+}$ &
(      3.1$_a$ &
        5.1$_b$ & 
        5.8$_c$ )\\
%
\botrule
\end{tabular}
&
\begin{tabular}[t]{cccc}
\toprule
\multicolumn{4}{c}{ZP}\\
  \colrule
  Mn$_{1}^{\rm B}$ &
(        8.4$_{a'}$ &
        0.5$_{b'}$ &
        4.8$_c$ )\\
 Mn$_{2}^{\rm B}$ &
(        9.5$_{a'}$ &
        0.5$_{b'}$ &
        4.3$_c$ )\\
Mn$_{1}^{\rm C}$ &
(        7.2$_{a'}$ &
         1.8$_{b'}$ &
         4.2$_c$ ) \\
Mn$_{2}^{\rm C}$ &
(        7.3$_{a'}$ &
         3.7$_{b'}$ &
         3.1$_c$ ) \\
  \botrule
\end{tabular}
\end{tabular}
%
%
%
%
\begin{tabular}{cc}
%
\begin{tabular}[t]{cccc}
\multicolumn{4}{l}{\pcmo}\\
\toprule
\multicolumn{4}{c}{CO1}\\
 \colrule
Mn$^{3+}$ & 
(         6.74$_x$ &
        0.89$_y$ &
         3.78$_c$ ) \\
Mn$^{4+}$ &
(         5.28$_x$ &
        3.94$_y$ & 
         3.45$_c$ )\\
%
\botrule
\end{tabular}
&
\begin{tabular}[t]{cccc}
\toprule
\multicolumn{4}{c}{CO2}\\
  \colrule
Mn$^{3+}$ &
(         6.29$_x$ &
         1.38$_y$ &
         3.71$_c$ ) \\
Mn$^{4+}_{\rm A}$ & 
(        4.67$_a$ &
        4.93$_b$ &
        3.49$_c$ ) \\
Mn$^{4+}_{\rm B}$ &     
(        4.22$_a$ &
        4.28$_b$ &
        3.90$_c$ )\\
%
  \botrule
\end{tabular}
\end{tabular}
\vspace{-0.5cm}
\end{table}
\bigskip

{\em Results: energetics and ferroelectricity. }
As shown in Tab.\ref{tbl:Ediff}, 
 the total energy in \pcm \: shows $E_{\rm CB}$$>$$E_{\rm ZP}$ (with the experimental structure), 
as consistent with previous DFT study.\cite{vincenzo} 
The calculated polarization (in Tab. \ref{tbl:Ediff}) shows a small $a$ component, $P_{a}$=$-$3.3$\mu$C/cm$^{2}$, and a large $b$ component, $P_{b}$=5.7$\mu$C/cm$^{2}$, originating from two different origins as reported in Ref.\cite{vincenzo}. 
The $P_{a}$ originates from the  dimerization of Mn ions along the orbital stripes (coexistence of bond- and site-centered CO mechanisms, as proposed in Ref.\cite{efremov}), consistent with the derivation from point charge model; 
$P^{\rm PCM}_{a}$=-2.1$\mu$C/cm$^{2}$.  
The dimerization  is reflected by the experimental structure: 
along the spin chains, the Mn-O-Mn bond angle is modulated by large and small angles, 
159.6$^{\circ}$ (intra-dimer Mn) and 154.8/157.8$^{\circ}$ (inter-dimer Mn). 
Considering that the large  Mn-O-Mn bond angle enhances the double-exchange coupling, 
we expect  the gravity center of charge to shift towards the middle of Mn-Mn dimers rather than being located on Mn sites,  resulting in a local electric dipole, suggested as the origin of improper ferroelectricity in La$_{0.5}$Ca$_{0.5}$MnO$_{3}$\cite{gianluca}.

On the other hand, the  $P_{b}$ 
component 
is due to artificially imposing the CE-AFM on top of the $P2_{1}nm$ crystal structure 
which leads $P_{b}$=0 in nonmagnetic configuration. 
When forcing the CE-AFM order, the different interaction between parallel/anti-parallel spin sites causes a Heisenberg-exchange-driven polarization of  purely electronic nature.

\begin{table}[h]
\vspace{-0.2cm}
\caption{
Total energy difference,  $\Delta_E^1 = E_{\rm ZP}-E_{\rm CB}$ and $\Delta_E^2 = E_{\rm CO2}-E_{\rm CO1}$ (meV/Mn) 
 for PCMO and \pcmo\ with different values of $U$ (eV). The polarization $P^{\rm tot}$ ($\mu$C/cm$^{2}$) is also reported. 
} 
\label{tbl:Ediff}
\begin{tabular}{|c|c|c|c|ccc|}
\hline 
\multicolumn{3}{|c|}{PCMO} \\
\hline
$U$ & $\Delta E^1$ & $P^{\rm tot}$ \\
\hline
0 & -51.7 &    (-3.56, 5.01, 0) \\
3 & -59.3 &    (-3.28, 5.66, 0) \\
5 & -50.9 &     (-3.18, 6.06, 0) \\
\hline
\end{tabular}
\begin{tabular}{|c|c|c|c|ccc|}
\hline
\multicolumn{3}{|c|}{\pcmo} \\
\hline
$U$ & $\Delta E^2$& $P^{\rm tot}$ \\
\hline
0 & -43.4 & (0, 2.18, 0)\\ 
3 & -47.2 & (0, 1.80, 0)\\
5 & -48.7 & (0, 1.55, 0)\\
\hline 
\end{tabular}
\vspace{-0.2cm}
\end{table}

{\em Ferroelectricity in bilayer manganites}. In  \pcmo, the calculated energy (Tab. \ref{tbl:Ediff}) shows that the
CO2 phase is energetically lower than CO1, irrespective of the chosen value of $U$. 
The calculated $P$ in CO2 phase has a non-zero and sizable $y$ component, $P^{\rm Berry}_{y}$=1.8$\mu$C/cm$^{2}$. 
As already pointed out in Ref.\cite{itoh}, the FE polarization is related to the stacking pattern of the CO bilayers. 
As shown in Fig. \ref{fig:COpattern}, the GdFeO$_{3}$-like tilting of MnO$_{6}$ octahedron largely displaces apical O (O$^{\rm ap}$) ions away from the bond center along the $b$ direction. 
One expects the displacement of O$^{\rm ap}$ ion to be suppressed by an elastic energy contribution, when Mn-$e_{g}$ electrons form strong Mn-O$^{\rm ap}$-Mn bonding along  the $c$ direction. 
Indeed, the Mn-O$^{\rm ap}$-Mn bond angles $\theta$, summarized in Tab. \ref{tbl:angle}, reveal 
$\theta^{44}<\theta^{33}$, i.e. O$^{\rm ap}$ ion is more displaced between Mn$^{4+}$ ions, with respect to the one between Mn$^{3+}$ ions.  The 
coupling between the  O$^{\rm ap}$ displacement and 
the staggered GdFeO$_{3}$-tilting pattern causes a local dipole inside the bilayer; the latter cancels out  with the nearby bilayer  in the CO1 phase (resulting in antiferroelectricity), whereas it gives rise to a net $P$ in the CO2 phase. 

Furthermore, we observe that the polarization value in CO2 is close to the $P$ calculated by using PCM 
$P^{\rm PCM}_{y}$(CD)=1.9 $\mu$C/cm$^{2}$, 
considering all Mn$^{3.5+}$ with homogeneous valence and 
$P^{\rm PCM}_{y}$(CO)=1.8 $\mu$C/cm$^{2}$, 
considering Mn$^{3+}$/Mn$^{4+}$ mixed valences. 
The almost identical value of $P^{\rm PCM}_{y}$(CD) and $P^{\rm PCM}_{y}$(CO)
shows that the origin of $\bm P$ lies in the apical O ion displacement, 
and is not crucially dependent on the Mn$^{3+}$/Mn$^{4+}$ CO pattern {\it per se}. In addition, this marks a clear difference
with respect to purely CO-induced ferroelectricity,  occurring for example in Fe$_{3}$O$_{4}$ and in iron-based fluorides.\cite{fe3o4_yamauchi, TTBletter}

\begin{table} \begin{center}
\caption{Mn-O$^{\rm ap}$-Mn angle $\theta$ ($^{\circ}$) in CD, CO1, and CO2 phases.} 
\label{tbl:angle}
\begin{tabular}{cccccc}
\toprule
            & $\theta^{33}$    & $\theta^{44}$  \\        
\colrule
CD  &    162.80       &  162.80          \\              
CO1    & 166.41     & 159.22              \\
CO2    & 162.91     &\    162.63(A), 159.28(B)       \\
\botrule
\end{tabular}
\end{center}
\vspace{-0.7cm}
\end{table}

Further insights can be gained by looking at the electronic structure. 
The OO both in the CO1 and CO2 phases can be clearly seen in the spin-density  plot reported in Fig.\ref{fig:chgsec}, 
showing a similarity between CO1 and ZP, and between CO2 and CB phase in PCMO, respectively (Fig.\ref{fig:chgsec}). 
In analogy with ZP, CO1 shows a dimerization between Mn ions so as to induce polarization in the MnO$_{2}$ plane; however,  the net induced polarization is canceled out by inter-layer antiferroelectric stacking. 
The difference of CO2 with respect to  CB is represented by two different Mn$^{4+}$ sites, namely Mn$^{4+}_{\rm A}$ and Mn$^{4+}_{\rm B}$, and by the anisotropic orbital shape.  
A careful look at the spin density plot reveals the anisotropy of the $e_{g}$ electron hopping. 
In the CO2 phase, the charge on the Mn$^{4+}_{\rm A}$ site shows a polar behaviour, slightly deviated from $x^{2}$-$y^{2}$ orbital shape, 
elongated and pointing toward one of the neighboring Mn$^{3+}$ sites. 
This is consistent with the dimerization of Mn ions, as present in the experimental structure, $i.e.$
 the larger angle of Mn$^{3+}$-O-Mn$^{4+}_{\rm A}$ than of Mn$^{3+}$-O-Mn$^{4+}_{\rm B}$ indicated in Fig. \ref{fig:chgsec}.  
Although the ZP state is realized thanks to the double-exchange nature of Mn $e_{g}$ electrons, 
here the dimerization seems to be induced by an external factor, {\em i.e.} the effective electric field induced by O$^{\rm ap}$ displacements. 
%
The spin-density plot in the $bc$ plane, shown in Fig. \ref{fig:chgsec}(e),  
reveals the lobes of Mn charge pointing along the $c$ direction, forming the inter-layer Mn-O-Mn bonding. 
Remarkably, in both CO phases, Mn ions form dimers so as to induce a local $\bm P^{\rm ZP}$ in MnO$_{2}$ planes, that cooperatively adds to the local $\bm P^{\rm ap}$ caused by the apical O$^{\rm ap}$ displacement, showing  an enhancement of the total $\bm P$. 
By using the calculated BEC tensors, the local contributions to $\bm P$ from each layer can be calculated separately and given by the summation of the product of the BEC ($Z_{}^{*}$) by the relevant ionic displacement. 
We therefore obtained $P_{y}$(CaO$^{\rm ap}$)=-0.36$\mu$C/cm$^{2}$, $P_{y}$(PrO$^{\rm ap}$)=$P^{\rm ap}$=0.65$\mu$C/cm$^{2}$, and $P_{y}$(MnO$^{\rm ip}_{2}$)=$P^{\rm ZP}$=0.71$\mu$C/cm$^{2}$ per layer, whereas 
the total $P_{b}$=1.8$\mu$C/cm$^{2}$. 
This notably shows that the polarization, $P^{\rm ap}$ and $P^{\rm ZP}$, originating from different mechanisms and different ions, give quantitatively similar contributions. 
The clear difference 
in the mechanism driving ferroelectricity between PCMO and \pcmo\ appear in $Z^{\rm d}$ (Tbl. \ref{tbl:bec.pcmo}):  
all Mn ions show a strongly {\em anisotropic} $Z^{\rm d}$ along dimers in the ZP phase, 
whereas {\em isotropic} $Z^{\rm d}$ of Mn$^{4+}$ is obtained in the CO2 phase. 
This implies that in \pcmo\ the $e_{g}$ double exchange mechanism is not relevant but rather structural effects play a role in the ferroelecricity. 

{\em Conclusions.}  We have investigated the ferroelectric half-doped manganites and  have made a comparison between pseudo-cubic PCMO and bilayer \pcmo. 
In PCMO, although there is an arbitrariness in the crystal structure (which was proposed to be either polar or not), a polarization is expected in the ZP structure. 
In bilayer \pcmo, the intercalation of layers gives rise to a peculiar situation, where both the Mn dimerization process and the apical oxygen displacements contribute to ferroelectric polarization, predicted to be of the order of several $\mu$C/cm$^2$. 
We finally note  BEC tensor is a good measure to quantify CO and OO in manganites. 

{\em Acknowledgments.} 
KY thanks Y. Taguchi, Y. Tokunaga and D. Okuyama for fruitful discussions 
and acknowledges kind hospitality at CNR-SPIN L'Aquila, where the manuscript was written. The research leading to these results received funding from the European Research Council under the European Community 7th Framework  Programme FP7 (2007-2013)/ERC Grant Agreement no. 203523 
and from JST-CREST ``Creation of Innovative Functions of Intelligent Materials on the Basis of the Element Strategy''. 
Some figures are plotted using the program VESTA\cite{vesta}. 


\end{document}